\documentclass[letterpaper, 10 pt, conference]{ieeeconf}
\usepackage{comment}
\usepackage{amssymb, amsmath,graphicx,charter, latexsym}
\usepackage{amsfonts}
\usepackage{amssymb, amsmath,graphicx,charter, latexsym}
\usepackage{graphicx}
\usepackage{epstopdf}
\usepackage{color}
\usepackage{amsfonts}
\usepackage{latexsym}
\usepackage{amssymb}
\usepackage{enumerate}
\usepackage{tikz,pgf,pgfplots}
\usepackage{cite}
\usepackage{bm}
\usepackage{bbm}
\usepackage{xcolor}
\usepackage{amsmath}
\newtheorem{lemma}{\quad Lemma}
\newtheorem{theorem}{\quad Theorem}

\usetikzlibrary{patterns}
\IEEEoverridecommandlockouts                              
\begin{document}
\title{Decentralized Throughput Maximizing Policies for Deadline-Constrained Wireless Networks
}

\author{Rahul Singh$^{1}$ and P. R. Kumar$^{1}$ 
\thanks{This material is based upon work partially supported by AFOSR Contract FA9550-13-1-0008,  and NSF under Contract Nos. CNS-1302182 and Science \& Technology Center Grant CCF-0939370.}
\thanks{$^{1}$Rahul Singh, and P. R. Kumar are with Texas A\&M University, College Station, TX 77840, USA.
        {\tt\small rsingh1@tamu.edu, prk@tamu.edu}}%
}
\maketitle
\IEEEpeerreviewmaketitle
\begin{abstract}
We consider multi-hop wireless networks serving multiple flows in which only packets that meet hard end-to-end deadline constraints are useful, i.e., if a packet is not delivered to its destination node by its deadline, it is dropped from the network. We design decentralized scheduling policies for such multi-hop networks that attain the maximum throughput of useful packets. The resulting policy is decentralized in the sense that in order to make a transmission decision, a node only needs to know the ``time-till-deadline" of the packets that are currently present at that node, and not the state of the entire network. The key to obtaining an easy-to-implement and highly decentralized policy is to replace the hard constraint on the number of simultaneous packet transmissions that can take place on the outgoing links of a node, by a time-average constraint on the number of transmissions. The policy thus obtained is guaranteed to provide maximum throughput. Analysis can be extended to the case of time-varying channel conditions in a straightforward manner.

Simulations showing significant improvement over existing policies for deadline based scheduling, such as Earliest Deadline First, and supporting the theory, are presented. 
\end{abstract}
\section{Introduction}
The focus of this paper is on multi-hop networks with per-packet end-to-end deadlines. Scheduling policies such as the back-pressure algorithm \cite{tassi1}, which are designed to achieve maximum throughput for traditional multi-hop wireless networks, provide guarantees only on average end-to-end delays, not per-packet delays. Due to the unreliable nature of the wireless medium, the per-packet delay along sample paths can become arbitrarily large. In-fact the back pressure algorithm or the MaxWeight scheduler has been shown to optimize the average end-to-end network delay only in the heavy-traffic regime \cite{sasha1}. However since the average delay grows roughly as $\frac{1}{1-\mbox{traffic intensity}}$, the delay becomes unbounded as traffic intensity approaches $1$. This is not desirable for applications such as cyber-physical systems where control-loops are closed over networks and system stability is sensitive to delays. 

In this paper, we consider multi-hop wireless networks serving various flows, in which if a packet is not delivered to its destination by its deadline, it is dropped from the network and not counted in the throughput. Since the wireless channel is unreliable, the outcome of packet transmissions is modeled as random processes. We suppose that each node can transmit multiple packets on its out-links. To incorporate an average power constraint, which presents itself as an average rate constraint on the out-going  communication channels linked to that node, we impose an upper bound on the average number of transmissions that can be made by a node in the network, a quantity which is allowed to depend on the individual node. Furthermore we assume that a node can transmit and receive packets on multiple channels, a property which can be achieved via various techniques such as TDMA, CDMA, OFDMA~\cite{ofdma1,ofdma2,ofdma3} which enable multiple access. We remark that such an assumption allows the network to make full use of the available resource-sharing disciplines.

The throughput of a flow is then the average number of packets delivered to its destination node per unit time, and our goal is to design decentralized scheduling policies that maximize the total throughput of the network. 

We use the scalarization principle \cite{hwang}, and consider a weighted throughput of all flows, whose maximum value will be one point on the Pareto frontier of the rate region. We pose the problem of obtaining a (weighted) throughput-maximizing scheduling policy as a Markov Decision Process (MDP). 

Our approach to solving this problem is via the Lagrangian dual of this MDP. The Lagrange multipliers associated with the rate-constraints are interpreted as prices paid by a packet to a node for transmitting its packet. 
 This renders our approach very different from that leading to the backpressure policy, where the Lagrange multipliers are associated with flows and not packets, and are therefore completely different, corresponding to queue lengths.

The resulting overall MDP decomposes conveniently into a ``unit-packet unit flow" MDP (Section~\ref{spsf}). This makes possible a decentralized packet-by-packet solution, where not just flows or nodes but individual packets can be individually optimized with respect to their treatment by nodes.
A node only needs to know the remaining lifetime till deadline of each packet that is present at the node and makes a decison on transmitting the packet based on that. Moreover the packet level MDP does not suffer from a severe curse of dimensionality since its state space is only the cardinality of the number of nodes multiplied by the relative deadline bound on the packet, and is thus relatively easily solved.
Thus, introduction of Lagrange multipliers, specifically prices for attempting packet transmissions, gives rise to a tractable and easy to implement decentralized scheduling policy. These Lagrange multipliers are shown to be computable in a decentralized online fashion. 

The key to our results is the long-term average constraint on the number of concurrent packet transmissions by a node. This can regarded as a long-term average relaxation of a more stringent hard constraint on the number of concurrent transmissions that a node may make at any time. This relaxation is in the same spirit as Whittle's relaxation for multi-armed bandits ~\cite{Whittle2011Book,Whi80}, where the constraint on the number of arms that can be simultaneously pulled is relaxed to a constraint on the average number of arms that can be pulled. Our relaxation therefore also has an asymptotic optimality property in the same manner that Whittle's relaxation has. In Whittle's case it is asymptotically optimal as the number of arms goes to infinity. Our relaxation is asymptotically optimal as the link capacities are scaled proportionally across the entire network, and results from the fact that on an average, not much is gained in terms of the total reward collection, by relaxing the hard constraint to an average constraint~\cite{weber}. The analysis can be carried out using the theory of large deviations for Markov processes, and it can be shown that the sub-optimality gap between the relaxed policy, and the optimal policy, is of the order $\frac{1}{\sqrt{\mbox{Network Capacity}}}$. Extensive simulations have been performed which show that the policy thus obtained drastically out-performs existing policies such as Back Pressure, Earliest Deadline First, and Debt-Based policies.

The system model considered excludes a key aspect of wireless networks, namely wireless interference. It could be incorporated by not allowing interfering links to schedule packet transmissions simultaneously. This would require the co ordinated effort of all the nodes in order to decide the optimal set of non-interfering links to be activated, and is the subject of future works.     
\section{System Model }\label{sm}
We consider wireless networks in which the data-packets have a hard deadline constraint on the time at which they are delivered to their destination nodes in order to be regarded as useful and counted in the throughput. The network comprises of several nodes $1,2,\ldots,V$ that are connected via wireless links. The wireless network is described by a directed graph in which there is a directed edge $i\to j$ if node $i$ can transmit packets to node $j$.
\begin{figure}[h]
	\centering
 \resizebox{5cm}{4cm}{
\begin{tikzpicture}
\node (a) at (0,0) [draw,circle,minimum size=.25cm] {$s_1$};
\node (b) at (1,0) [draw,circle,minimum size=.25cm] {$b$};
\node (c) at (2,0) [draw,circle,minimum size=.25cm] {$d_1$};
\node (d) at (0,1) [draw,circle,minimum size=.25cm] {$s_2$};
\node (e) at (0,-1) [draw,circle,minimum size=.25cm] {$d_2$};
\node (f) at (1,1) [draw,circle,minimum size=.25cm] {$s_3$};
\node (g) at (1,-1) [draw,circle,minimum size=.25cm] {$d_3$};

\draw [->] (a) -- (b);
\draw [->] (b) -- (c);
\draw [->] (d) -- (a);
\draw [->] (a) -- (e);
\draw [->] (f) -- (b);
\draw [->] (b) -- (g);

\end{tikzpicture}}
\caption{Multi-hop network serving three flows $f_1$ on the route $s_1\to b \to d_1$, $f_2$ on the route $s_2\to s_1\to d_2$, and $f_3$ on the route $s_3 \to b \to d_3$.}
\label{fig2}
\end{figure}
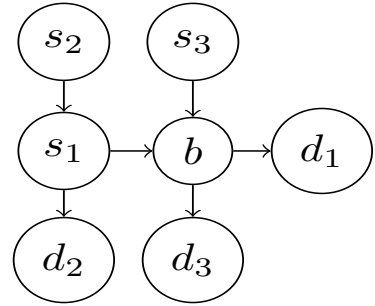

We assume that time is discrete, and evolves over discrete time slots numbered $1,2,\ldots$. One time-slot is the time taken to attempt a packet transmission over any link in the wireless network. The network is shared by $F$ flows $f_1,f_2,\ldots,f_F$. Each flow $f_i$ has a pre-determined route comprised of a sequence of links that connect the source node of the flow $f_i$ to its destination node; see Figure~\ref{fig2}. 

The wireless channel between any two nodes is allowed to be random. If a packet of any flow $f$ is attempted on the wireless link $l$, then the transmission is successful with probability $p_l$. The outcomes of packet transmission attempts are independent across links and time-slots. 

Each node $v$ has an average rate-constraint $ M_v$, which is the maximum number of packet transmission attempts per unit time that it can make. We will use the notation $l\in v$ to mean that link $l$ is in the set of out-links from node $v$, and thus $v$ can use the link $l$ for transmitting packets. Let $u^l_f(t)$ be the random variable which denotes the number of packets of flow $f$ that are attempted for transmission on link $l$ at time $t$. The rate constraints are given by,
\begin{align}\label{shannon}
&\limsup_{T\to\infty}\mathbb{E}\left(\frac{\sum_{t=1}^{T}\sum_{f}\sum_{l\in v}u_f^l(t)}{T}\right) \leq M_v,\\
&\qquad  \forall v\in \{1,2,\ldots,V\}\notag.
\end{align} 
Note that we allow a node to transmit and receive packets simultaneously over several outgoing links, which is possible via various techniques such as TDMA, OFDMA, CDMA etc.\cite{ofdma1,ofdma2,ofdma3}, and hence $M_v$ can be larger than $1$. The rate-constraint is due to the fact that wireless nodes have power constraints, which in turn induces constraints on communication.

Each packet that is generated by the network has a ``relative-deadline", and if the packet is not delivered to its destination within this deadline, it is dropped from the network and will not be transmitted in future time-slots. More precisely, if a packet has a relative-deadline $\delta$, and is generated at the beginning of time-slot $\tau$, then either it is delivered to its destination node by time-slot $\tau + \delta$, or it is discarded from the network.

We assume that the relative deadlines of the packets are i.i.d. and bounded by a fixed $\Delta$, where the distribution of the relative deadlines depends on the flow that the packet belongs to. The relative deadline becomes known to the network as soon as the packet is generated.

Our analysis can be extended in a straight-forward manner to consider the case when the relative-deadline of a packet is an arbitrary stochastic process that is adapted, though we eschew that here for brevity. When so chosen as an adapted stochastic process, some very useful models can be captured. For example, suppose we have a video context, and we have a frame buffer at the receiver, then the relative deadline is equal to the ``remaining play time" left in the frame buffer since we don't want the buffer emptied. In that case Relative Deadline $= -$ (Elapsed time since the Last time that Destination Buffer was empty, i.e., the current age of the ``busy epoch") + (Number of packets that arrived at the Destination since then )$ \times $ (Time to play one packet). Note that in this case the deadline process depends on the policy being used.

We assume that the inter-arrival times of packets at different source nodes for each flow are governed by renewal processes having finite means. The throughput attained by a flow $f$ under a policy is defined to be, 
\begin{align}\label{tp}
q_f :=\liminf_{T\to\infty} \mathbb{E}\left(\frac{\sum_{t=1}^{T}d_f(t)}{T}\right), 
\end{align} 
where the random variable $d_f(t)$ is $=1$ if a packet of flow $f$ is delivered to its destination at time $t$, and is $=0$ otherwise, and expectation is under the policy that is applied. A throughput vector $\boldsymbol{q}$ that can be achieved via some scheduling policy will be called an ``achievable throughput vector", the set of all achievable throughput vectors constitutes the ``rate-region", and a scheduling policy that achieves the rate-region is said to be throughput-optimal. 

Of course, all the above definitions depend on the random process which decides the relative-deadlines of the packets, and so therefore does the rate-region. Thus we might call such networks as ``deadline-constrained networks".  

Vectors will be in bold font, and by $\mathbb{R}^{N}_{+}$ we refer to $N$ dimensional vectors which are non-negative component wise.
\section{Previous Works}\label{pw}
References~\cite{hbk,hou1,hou2} consider a network model in which multiple flows share a single-hop network, and all the packets across every flow have the same relative deadline. Restricting consideration to a periodic arrival process with relative deadline equal to the period at any given time at most, only one packet belonging to each flow is present in the network. References~\cite{modiano,jagan} consider a similar one-hop network model and characterize the throughput maximizing policy. Clearly the single-hop model is restrictive. 

Reference~\cite{atilla1} considers the challenging problem of scheduling deadline-constrained packets over a multi-hop network, but the proposed policies are not shown to have any provable guarantees on the resulting  throughput. To the best of the author's knowledge,~\cite{shroff1} is the only work which provides a provable sub-optimal policy for deadline-constrained networks, though it only concerns wired networks. Moreover the policies proposed in~\cite{shroff1} guarantee only a fraction 
\begin{align*}\frac{1}{\mbox{length of the longest route in network}}\end{align*} of the maximum possible throughput, i.e., only a small fraction of the capacity region. 

Though the MaxWeight policy \cite{sasha1} has been popular among researchers in designing multi-hop scheduling-routing policies such as the Back-Pressure \cite{tassi1}, or decentralized CSMA wireless scheduling policies \cite{wlrnd1,srikant1}, it should be noted that an application of the MaxWeight principle to design multi-hop scheduling policies for wireless networks results in the number of possible switch vectors available to the scheduler \cite{sasha1} growing roughly exponentially in the quantities: total number of flows, sum of the relative deadlines of flows, and the sum of the route lengths of flows. While some of the previous works on deadline-constrained scheduling \cite{hbk,hou1,hou2} have been successful in designing deadline-constrained throughput optimal ``debt-based" policies for single-hop networks, any generalization to the case of multi-hop wireless network seems difficult. 

Joint routing scheduling policies designed for multi hop networks, such as the Backpressure~\cite{tassi1} are throughput optimal, but perform poorly with respect to the end-to-end delay~\cite{d1,d2,d3,d4}. References~\cite{rahul} and~\cite{rahul1} perform a fluctuation analysis of the ``packet starvation" that occurs in deadline constrained networks, while \cite{Rahul2015,guosingh,rs} develop a framework for service regularity for networks shared by multiple flows. 

Thus, presently the knowledge of provably optimal scheduling policies for deadline-constrained multi-hop wireless networks is severely limited.
Any optimal scheduling policy needs to take into account how much time each packet has spent in the network, as well as the channel reliabilities of the links that the packets have to traverse in order to reach the destination node.
\section{Characterizing the Rate Region}\label{rr}
The rate-region of the network (defined in Section~\ref{sm}) will be denoted by $\boldsymbol{\Lambda}$ . We note that in order to characterize it, it is sufficient to characterize the set of Pareto-optimal vectors $\mathbf{q}\in \boldsymbol{\Lambda}$, defined as
\begin{align*}
\left\{\mathbf{q}: \exists \boldsymbol{\alpha}\in \mathbb{R}^{N}_{+} \mbox{ such that }\mathbf{q}\in \arg\max_{\mathbf{y}\in \Lambda} \sum_{f}\alpha_f y_f     \right\},
\end{align*}
since $\boldsymbol{\Lambda}$ is simply its closed convex hull. Thus the problem of obtaining the set $\boldsymbol{\Lambda}$ is equivalent to that of finding scheduling policies which maximize a non-negatively weighted sum of throughputs.
\subsection{Constrained MDP Formulation}
The problem of maximizing a non-negatively weighted sum of throughputs subject to rate-constraints can be posed as a Constrained Markov Decision Process (CMDP)~\cite{altman1}. The system state can be described as follows. At time $t$, each packet in the network belonging to flow $f$ is described by the two tuple $\left(l,s\right)$, where $l$ is the link at which the packet is present, and $s$ is the time-to-go till its deadline. The state of the network is then simply given by the states of each packet present in the network.

Since the number of packets in the network at any time is bounded (because the relative deadlines are bounded only a bounded number of packets can hope to be served in this time period and others can be safely dropped), the system state $x(t)$ takes on finitely many values. A scheduling policy $\pi$ has to choose, for each time $t$, at each node, which packets to transmit from the set of packets available to it. The probability distribution of the system state at time $t+1$, $x(t+1)$, depends only on the system state at time $t$, $x(t)$ and the action chosen at time $t$ by the policy $\pi$. The problem of maximizing the throughput subject to node-capacity constraints~\eqref{shannon} is posed as a Constrained Markov Decision Process, where a reward of $\alpha_f$ is received when a packet of flow $f$ is delivered to its destination.

Thus a policy maximizing the network throughput solves the following optimization problem:
\begin{align}\tag{Primal MDP}
&\max_{\pi} \liminf_{T\to\infty}\frac{1}{T}\mathbb{E}\left(\sum_{f}\sum_{t=1}^{T}\alpha_i d_f(t)\right)\mbox{, such that } \notag\\ 
&\limsup_{T\to\infty}\frac{1}{T}\mathbb{E}\left(\sum_{t=1}^{T}\sum_{f}\sum_{l\in v}u_f^l(t)\right) \leq M_v, \forall v\in \{1,2,\ldots,V\}.
\label{op}
\end{align}
We note that the above CMDP, parameterized by the vector $\boldsymbol{\alpha}:=\left(\alpha_1,\ldots,\alpha_F\right)$ is solved by a Stationary Randomized Policy (\cite{altman1}). Since the state-space of the network, and the number of link-capacity constraints~\eqref{shannon} are finite, it follows that there is a finite set $\{\pi_1,\pi_2,\ldots,\pi_M\}$ of Stationary Randomized Policies such that for each value of $\boldsymbol{\alpha}$, there is a policy that belongs to this set and solves the CMDP~\eqref{op} (\cite{altman1}). Let $\boldsymbol{\gamma}_1,\boldsymbol{\gamma}_2,\ldots,\boldsymbol{\gamma}_M$ be the vectors of throughputs associated with the policies $\pi_1,\pi_2,\ldots,\pi_M$. We then have the following characterization of $\boldsymbol{\Lambda}$.
\begin{lemma}\label{lemma1}
\begin{align*}
\boldsymbol{\Lambda} = \left\{\mathbf{q}: \mathbf{q}=\sum_{i=1}^{M} \boldsymbol{\gamma}_i c_i, c_i\geq 0,\sum_i c_i \leq 1    \right\}.
\end{align*}
\end{lemma}

Note that the number of Stationary Markov policies is exponentially large in the parameter: \\maximum possible number of packets in the network $\times$ maximum path length of the flows $\times$ maximum possible relative deadline.

Hence using Lemma~\ref{lemma1} to compute $\boldsymbol{\Lambda}$ is out of the question. Thus we seek to design low-complexity decentralized scheduling policies that achieve the region $\boldsymbol{\Lambda}$.

\section{The Dual MDP}\label{a1}
We write the Lagrangian for the Primal MDP (\ref{op}) as,
\begin{align}\label{lang}
&\mathcal{L}(\pi,\boldsymbol{\hat{\lambda}})=\notag\\
& \liminf_{T\to\infty}\frac{1}{T}\mathbb{E}\left(\sum_{f}\sum_{t}\alpha_f d_f(t)\right)\notag\\
&+\sum_{v} \hat{\lambda}_{v}  \left( \liminf_{T\to\infty}\frac{1}{T}\mathbb{E}\left(\sum_{t=1}^{T}\sum_{f}\sum_{l\in v} \left[-u_f^l(t)\right]\right) \right)\\
& +\sum_{v} \hat{\lambda}_{v}M_v \notag.
\end{align}
The corresponding dual is,
\begin{align}\label{dual}
D(\boldsymbol{\hat{\lambda}}) = \max_{\pi} \mathcal{L}(\pi,\boldsymbol{\hat{\lambda}}).
\end{align}

Next we develop a useful upper-bound on $D(\boldsymbol{\hat{\lambda}})$, the proof of which follows from the sub-addivity of $\liminf$ operation and super-additivity of $\limsup$ operation.
\begin{lemma}\label{l1}
\begin{align}\label{decom}
D(\boldsymbol{\hat{\lambda}})\leq \sum_f V_f(\boldsymbol{\hat{\lambda}})+\sum_{v} \hat{\lambda}_{v}M_v,
\end{align}
where 
\begin{align*}
& V_f(\boldsymbol{\hat{\lambda}}) := \\
&\max_{\pi_f}\limsup_{T\to\infty}\frac{1}{T}\mathbb{E}\left( \sum_{t}\left(\alpha_f d_f(t)+\sum_v \sum_{l\in v}\hat{\lambda}_{v}  \left[-u_f^l(t)\right]\right)\right),
\end{align*}
and $\pi_f$ is a policy that schedules packets of flow $f$.
\end{lemma}
\section{The Single Flow, Single Packet Problem}\label{spsf}
Consider the expression in the r.h.s. of~\eqref{decom}. We note that the introduction of the vector $\boldsymbol{\hat{\lambda}}$ while considering the dual problem~\eqref{dual} decouples the problem~\eqref{op} into $F$ number of \emph{single-flow problems}, wherein if a packet of flow $f$ is present at node $v$ at time $t$, and it is attempted on a link $l$ that belongs to the set of out going links of node $v$, then flow $f$ is charged a price of amount $\lambda_{l}$. Thus $\hat{\lambda}_v$ can be interpreted as the price that the node $v$ charges for each use of any of its out-links. 

 We note that the above Lagrange multipliers are very different from the Lagrange multipliers employed in deriving the Backpressure policy. There the Lagrange multipliers correspond to queue lengths, whereas here, as we will show below, they are the price paid by a packet to a node for the privilege of being transmitted. It is this difference that will allow us to decompose the overall problem on packet-by-packet basis. It is this that will allow us to obtain a decentralized policy that is not only decentralized across flows, but also over nodes, and in fact over packets, with not even any coupling between packets at the same node.
 
We will first solve these $F$ single-flow problems. Then we will show that the policy that implements the solution to the single-flow problem for each flow solves the relaxed problem. Denote by $\pi_f$ a policy for flow $f$. The policy $\pi_f$ knows the value of the node prices $\hat{\lambda}_v$ of all the nodes $v$ that lie on the route of flow $f$, and the state of all the packets in the network that belong to flow $f$, and has to make a decision whether or not to schedule a packet transmission for flow $f$ at a link $l$, but does not need to keep track of the state of other flows.

The single-flow problem, parametrized by the vector of prices $\boldsymbol{\hat{\lambda}}$ is to find the policy $\pi_f$ that maximizes,
\begin{align}
V_f(\boldsymbol{\hat{\lambda}})&= \max_{\pi_f}\limsup_{T\to\infty}\frac{1}{T}\mathbb{E}\sum_{j=1}^{N_f (t)}\left( \sum_{t}\left(\alpha_f d^j_f(t)+\right.\right.\notag\\
&\qquad\qquad \qquad \left.\left.\sum_v \sum_{l\in v}\hat{\lambda}_{v}  \left[-u_f^{l,j}(t)\right]\right)\right),
\label{sfp}
\end{align}
where $u_f^{l,j}(t)$ assumes the value $1$ if the $j$-th packet belonging to the flow $f$ is served at link $l$ in time-slot $t$, and $0$ otherwise, $d^j_f(t)$ assumes the value $1$ if the $j$-th packet of flow $f$ is delivered to its destination at time $t$, and $N_f(t)$ is the number of packets for flow $f$ that arrive into the network by time $t$. Also note that $u_f^{l,j}(t)$ can assume the value $1$ only if the $j$-th  packet of flow $f$ is present at link $l$ at time $t$. 

Note that while introducing the single-flow problem, we have reduced it to a \emph{single-flow single-packet problem}, i.e. the total reward earned under the application of a policy $\pi_f$ is the sum of the rewards it earns from each packet. \emph{Thus the policy that solves the single-slow problem makes a decision whether or not to schedule a packet for flow $f$ depending only on the state of that single packet!} Thus the introduction of node-prices $\boldsymbol{\hat{\lambda}}$ not only decouples the original problem into $F$ separate problems, but further separates the problem for a single flow into that involving only a particular packet of that flow. 

Next we describe and solve the \emph{single-packet-single-flow problem}. We note that the state of a packet is much smaller than the state of the entire network. The state of a packet only consists of the node it is at, and its time till deadline. Therefore the Markov Decision Process for the single packet problem is much more tractable than the MDP for the overall system. 
To make the discussion simpler, since we are dealing with a single-flow, we will assume that the nodes have been re-labelled so that the source node is labelled as $1$, while the destination node is $L+1$. With this notation in place, the link between nodes $i$ and $i+1$ is labelled as $l_i$ (so that $l_i$ is the $i$-th link on the route). The single flow under consideration has to be served using the route $l_1,l_2,\ldots,l_L$. A single packet for this flow is generated at time $t=1$ at the source node $1$, and needs to be delivered to the node $L+1$ by the time $D+1$. The wireless link (channel) $l$ has a channel reliability $p_l$. Thus the packet has a relative deadline of $D$ time-slots, and has to traverse $L$ hops in order to reach the destination.

If the packet is delivered in a time-slot $t\in \{1,2,\ldots,D+1\}$, then a reward of $\alpha$ is received and the packet leaves the system. However if the packet is not delivered by the time $D+1$, it is removed from the system without generating a reward of $\alpha$ units. If the packet is present at node $v$ and it is attempted, then it is charged an instantaneous price of $\hat{\lambda}_v$. The total reward accrued by a packet is any rewards it obtains when it reaches its destination on time, minus all the prices it paid at all the links along its path to the destination.

The policy $\pi_f$ has to decide at each time whether or not to schedule the packet's transmission, so as to maximize the total expected reward earned in the time $\{1,2,\ldots,D+1\}$.

The single-packet single-flow problem is clearly a finite-state stochastic dynamic programming problem wherein the state of the packet assumes values in
\begin{align*}
\{\left(l_i,s\right): 0\leq i\leq L \mbox{ and } L-i\leq s\leq D-i\},
\end{align*}
where $l_i$ is the link at which the packet is present and $s$ is the ``time remaining till deadline". 

Let $V^{\boldsymbol{\hat{\lambda}}}(l,s)$ be the maximum expected reward that can be earned by the packet starting in state $\left(l,s\right)$. The function $V^{\boldsymbol{\hat{\lambda}}}(l,s)$ is not to be confused with the function $V_f(\boldsymbol{\hat{\lambda}})$ defined for the single-flow problem in~\eqref{sfp}. 
 The super scipt shows the dependence of the value function $V^{\boldsymbol{\hat{\lambda}}}(l,s)$ on the price vector $\boldsymbol{\hat{\lambda}}$. At times, the superscript $\boldsymbol{\hat{\lambda}}$ will be omitted to make the notation simpler.

Clearly $V(l_i,L-i)=0$ for $i=1,2,\ldots,L$, since a packet at link $l_i$ needs at least $L-i$ time-slots to reach the destination node, but since the time-to-go till deadline is $L-i$, it is dropped from the network. Using the principle of optimality, the values $V(l,s)$ are computed as,
\begin{align}\label{dp}
V(l_i,s) & = \max \{ V(l_i,s-1),\notag\\
& - \lambda_{l_i} + p_{l_i} V(l_{i+1},s-1) +\left(1-p_{l_i}\right) V(l_i,s-1) \},
\end{align}
where the second expression within the braces corresponds to the rewards associated with scheduling the packet transmission in state $(l_i,s)$. The first expression is the reward under the choice of not scheduling, and the optimal action is the one that achieves the maximum on the rhs. The computations in~\eqref{dp} are performed for all states in the order \\
$\left(l_L,1)\right),\left(l_L,2\right),\ldots,\left(l_{L-1},2\right),\ldots$, and thus the optimal actions and value function are obtained in $LD$ steps.

Let $\phi_f(l,s)=0$ if the first term in the rhs of~\eqref{dp} is strictly larger, and $=1$ if the second term  is strictly larger. For values of states $(l,s)$ in which both the terms in the rhs of~\eqref{dp} are equal, set $\phi_f(l,s)$ to be any value in the set $\left[0,1\right]$. Let $\boldsymbol{\phi}_f$ be the vector consisting of the values $\phi_f(l,s)$. Denote by $\pi_f(\boldsymbol{\hat{\lambda}},\boldsymbol{\phi}_f)$ the policy which transmits the packet of flow $f$ with a probability $\phi_f(l,s)$ when the packet is in state $(l,s)$. By construction,  $\pi_f(\boldsymbol{\hat{\lambda}},\boldsymbol{\phi}_f)$ solves the single-flow-single-packet problem.

Thus the optimal policy $\pi_f(\boldsymbol{\hat{\lambda}},\boldsymbol{\phi}_f)$ can be parameterized by the vector $\boldsymbol{\phi}_f$. Since the total rewards earned in the Single Flow Problem~\eqref{sfp} is the sum of rewards earned from each packet, we have,  
\begin{lemma}\label{l3}
Policies $\pi_f(\boldsymbol{\hat{\lambda}},\boldsymbol{\phi_f})$ solve the Single Flow Problem~\eqref{sfp}.
\end{lemma}
It should be noted that the optimal policy depends on the price vector $\boldsymbol{\hat{\lambda}}$.
The key to the dramatic decomposition of an originally very complex problem
of optimizing the behavior of the entire network lies in the fundamental nature of the constraint: It is an average constraint on number of packets. One can regard this as a relaxation of a hard constraint on the number of concurrent packet transmissions allowed to a node, a relaxation that has an asymptotic optimality property. To see this, an analogy with the multi-arm bandit problem where one is allowed up to $n$ simultaneous pulls per play is revealing. For this problem, there is no Gittins Index; however Whittle~\cite{Whi80} has shown that if this constraint is relaxed to an average constraint of no more than $n$ pulls, then one can obtain a decomposition. Moreover this constraint is optimal in the limit as the number of arms of different types and arm pulls are increased in proportion to infinity. In our case too, as the link capacities are scaled proportionally across the entire network we get asymptotic optimality.
\section{Solution of Primal MDP}\label{pmdp}
Given price vector $\boldsymbol{\hat{\lambda}}$ and vector $\boldsymbol{\phi}=\{\boldsymbol{\phi_f}, f=1,2,\ldots,F\}$, let us denote by $\pi(\boldsymbol{\hat{\lambda}},\boldsymbol{\phi})$ the policy that jointly follows the rule $\pi_f(\boldsymbol{\hat{\lambda}},\boldsymbol{\phi_f})$ for each flow $f$, where $\pi_f(\boldsymbol{\hat{\lambda}},\boldsymbol{\phi_f})$ is as defined in Section~\ref{spsf}. First, we show that there exist values of the vectors $\boldsymbol{\hat{\lambda}},\boldsymbol{\phi}$ such that the policy $\pi_f(\boldsymbol{\hat{\lambda}},\boldsymbol{\phi_f})$ solves the Primal MDP.
 \begin{lemma}\label{l4}
$D(\boldsymbol{\hat{\lambda}}) = \mathcal{L}(\pi(\boldsymbol{\hat{\lambda}},\boldsymbol{\phi}),\boldsymbol{\hat{\lambda}})$.
\end{lemma}
\proof
From the definition of the dual function~\eqref{dual} and the function $V_f$~\eqref{sfp}, it follows that,
\begin{align*}
\mathcal{L}(\pi,\boldsymbol{\hat{\lambda}} ) \leq D(\boldsymbol{\hat{\lambda}}) \leq \sum_f V_f +\sum_{v} \hat{\lambda}_{v}M_v +\sum_{v} \hat{\lambda}_{v}M_v,
\end{align*}
for any policy $\pi$. However since the policies $\pi(\boldsymbol{\hat{\lambda}},\boldsymbol{\phi})$ are Stationary Randomized Markov policies (all the $\liminf$s and $\limsup$s in the definition of its Lagrangian and also the corresponding rewards associated with the single-flow problem change to $\lim$), it follows from Lemma~\ref{l3} that $\sum_f V_f$ is the value of $\mathcal{L}(\pi,\boldsymbol{\hat{\lambda}} )$ evaluated at the policies $\pi(\boldsymbol{\hat{\lambda}},\boldsymbol{\phi})$, and thus the inequality above becomes an equality,
\begin{align}\label{dual1}
\mathcal{L}(\pi(\boldsymbol{\hat{\lambda}},\boldsymbol{\phi}),\boldsymbol{\hat{\lambda}} )= D(\boldsymbol{\hat{\lambda}}) = \sum_f V_f +\sum_{v} \hat{\lambda}_{v}M_v.
\end{align}

\begin{theorem}\label{duality}
Consider the Primal MDP~\eqref{op} and its associated dual problem defined in~\eqref{dual}. There exists a price-vector $\boldsymbol{\hat{\lambda}}^{\star}$, and vectors $\boldsymbol{\phi_f}^{\star}, f=1,2,\ldots,F$, such that $(\pi(\boldsymbol{\hat{\lambda}}^{\star},\boldsymbol{\phi}^{\star}),\boldsymbol{\hat{\lambda}}^{\star})$ is an optimal primal-dual pair, and thus the policy $\pi(\boldsymbol{\hat{\lambda}}^{\star},\boldsymbol{\phi}^{\star})$ solves the relaxed problem.
\end{theorem}
\proof
We use the ergodic control approach developed in~\cite{ergodic1,ergodic2,altman1}, particularly useful for constrained MDPs, wherein an average-cost MDP can be viewed as a linear programming problem. More precisely, the infinite-horizon average-reward problem can be posed as that of optimizing a linear cost function over a convex set after one considers occupation measures of the combined state-cum-control Markov process. If $\nu^{\pi}(\cdot)$ is the occupation measure induced under the policy $\pi$, the problem reduces to maximizing 
\begin{align*}
\sum_{\mbox{state,control}} \mbox{reward(state,control)}\nu^{\pi}(\mbox{state,control}).
\end{align*} 
We further note that the constraints in the  Primal MDP are linear functions of $\nu(\cdot)$, and hence the set over which optimization is to be carried out is convex.

Thus if we show that there exists a policy such that the constraints in the Primal MDP~\eqref{op} hold with strict inequality ``$>$" rather than ``$\geq$", then the proof of the Lemma would follow from Slater's condition~\cite{bkas}. But then the policy which never schedules any packet regardless of the value of the system state satisfies the constraints with strict $>$ ( the parameters $M_v$ are assumed to be $>0$, i.e. the total capacity of all the out-links associated with each node $v$ is assumed to be $>0$.)

We note that the policy $\pi(\boldsymbol{\hat{\lambda}}^{\star},\boldsymbol{\phi}^{\star})$ is decentralized, and each node only needs the knowledge of the time-till-deadlines of the packets it has. 
\subsection{Obtaining \large $\boldsymbol{\hat{\lambda}}^{\star},\boldsymbol{\phi}^{\star}$}
The policy obtained above is decentralized and easy to implement; one has only to obtain the price vector $\boldsymbol{\hat{\lambda}}^{\star}$ that solves the dual problem~\eqref{dual}. Next we provide an iterative algorithm to find $\boldsymbol{\hat{\lambda}}^{\star}$ that can be implemented in a decentralized manner, i.e., the nodes need only local information of the total link usage on all of its out-links. Thus the nodes need not communicate amongst themselves to obtain global information about the state of the network in order to derive $\boldsymbol{\hat{\lambda}}^{\star}$, which rules out involvement of any communication over heads.

Since the dual function $D(\boldsymbol{\hat{\lambda}})$ is convex, we can use the sub-gradient method \cite{shor} to iteratively find  $\boldsymbol{\hat{\lambda}}^{\star}$. From Lemma~\ref{l4}, we have $D(\boldsymbol{\hat{\lambda}}) = \mathcal{L}(\pi(\boldsymbol{\hat{\lambda}},\boldsymbol{\phi}),\boldsymbol{\hat{\lambda}})+\sum_{v} \hat{\lambda}_{v}M_v$, where the value of the Lagrangian $\mathcal{L}(\pi(\boldsymbol{\hat{\lambda}},\boldsymbol{\phi}),\boldsymbol{\hat{\lambda}})$ does not depend on the probability with which the packet is scheduled for values of the states in which the active (scheduling) and passive (not scheduling) actions yield equal rewards. Thus in order to find the value of dual function $D(\boldsymbol{\hat{\lambda}})$, it suffices to fix the randomizing probability to be $0$ for any value of the state in which the active and passive actions yield equal rewards. We can choose as sub-gradients the gradient associated with the policy $\pi(\boldsymbol{\hat{\lambda}},\boldsymbol{\phi})$, i.e., the quantities $\frac{\partial \mathcal{L}(\pi(\boldsymbol{\hat{\lambda}},\boldsymbol{\phi}),\boldsymbol{\hat{\lambda}})}{\partial \hat{\lambda}_v}, v =1,2,\ldots,V$.
Next we provide the expression for the gradients $\frac{\partial \mathcal{L}(\pi(\boldsymbol{\hat{\lambda}},\boldsymbol{\phi}),\boldsymbol{\hat{\lambda}})}{\partial \hat{\lambda}_v}, v =1,2,\ldots,V$, which helps us in interpreting it as congestion at the nodes, and the node-prices as a means to control the congestion.

From the description of the Single Flow Problem~\eqref{sfp}, it follows that the total cost a packet pays due to the link-price  $\hat{\lambda}_v$ is equal to $\hat{\lambda}_v$ times the total number of times that it uses an out-link at node $v$. Thus letting $\tilde{\tau}(f,v)$ to be a random variable that has the same distribution as the total link-usage per-unit time-slot by packets of flow $f$ at node $v$, we have from~\eqref{decom},
\begin{align}\label{partial1}
\frac{\partial D}{\partial \hat{\lambda}_v} =M_v- \mathbb{E}_{\pi(\boldsymbol{\hat{\lambda}},\boldsymbol{\phi})} \sum_f \tilde{\tau}(f,v).
\end{align} 
The quantity $\mathbb{E}_{\pi(\boldsymbol{\hat{\lambda}},\boldsymbol{\phi})} \sum_f \tilde{\tau}(f,v)$ is the total link-usage at node $v$, and thus measures the congestion at node $v$. The price $\hat{\lambda}_v$ is thus a means to prevent congestion at node $v$.

The iterations are given by
\begin{align}\label{partial}
\boldsymbol{\hat{\lambda}}^{k+1} = \boldsymbol{\hat{\lambda}}^{k} - \alpha_k \boldsymbol{g}^{k},
\end{align}
where $\boldsymbol{g}^{k}$ is the sub-gradient evaluated at $\boldsymbol{\hat{\lambda}}^{k}$ as in ~\eqref{partial1}. Since the value of the $i$-th component of sub-gradient as provided in equation~\eqref{partial1} is a local information, meaning that node $i$ can calculate its total link-usage without resorting to any communication with other nodes, these iterations can be performed locally.
 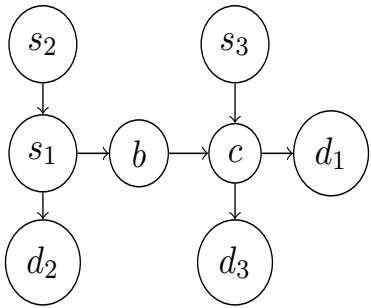
\begin{figure}[h]
 \centering
\resizebox{5cm}{4cm}{
\begin{tikzpicture}
\node (a) at (0,0) [draw,circle,minimum size=.25cm] {$s_1$};
\node (b) at (1,0) [draw,circle,minimum size=.25cm] {$b$};
\node (c) at (2+1,0) [draw,circle,minimum size=.25cm] {$d_1$};
\node (h) at (2,0) [draw,circle,minimum size=.25cm] {$c$};
\node (d) at (0,1) [draw,circle,minimum size=.25cm] {$s_2$};
\node (e) at (0,-1) [draw,circle,minimum size=.25cm] {$d_2$};
\node (f) at (2,1) [draw,circle,minimum size=.25cm] {$s_3$};
\node (g) at (2,-1) [draw,circle,minimum size=.25cm] {$d_3$};

\draw [->] (a) -- (b);
\draw [->] (b) -- (h);
\draw[->] (h)--(c);
\draw [->] (d) -- (a);
\draw [->] (a) -- (e);
\draw [->] (f) -- (h);
\draw [->] (h) -- (g);

\end{tikzpicture}}
\caption{Multi-hop network serving three flows $f_1$ on the route $s_1\to b \to c \to d_1 $, $f_2$ on the route $s_2\to s_1\to d_2$, and $f_3$ on the route $s_3\to c\to d_3$.}
\label{fig3}
\end{figure}
The analysis here can be extended to cover the case of time-varying channels so as to incorporate the wireless fading. This is accomplished by appending the state with the state of the wireless channels across the network.
 \section{Simulations}
Apart from the scheduling policy deployed, we note that for a deadline-constrained network, the net throughput depends upon the following factors: relative deadlines of the packets, the random process that decides the number of the packet arrivals in different flows, the channel reliabilities of various links and the node capacities $M_v$'s.

We compare the optimal policy derived in the paper with the policy that implements the EDF scheduling rule at each node in the network. The EDF policy is known to be optimal in the case of single-hop network.

\begin{figure}[!t]
	\centering
	\includegraphics[width=0.5\textwidth]{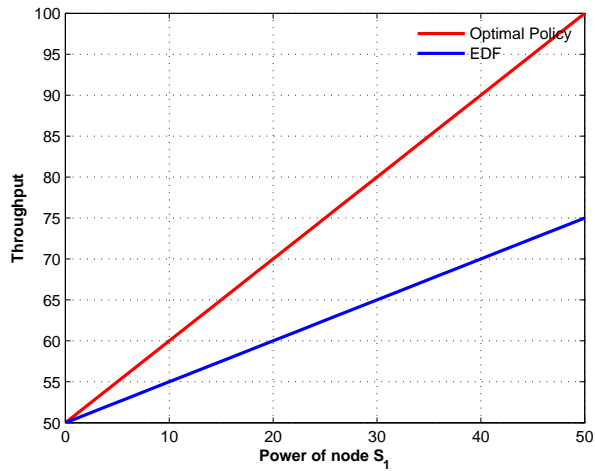}
	\caption{Plot showing the throughputs attained by the policies for the network of Figure~\ref{fig3} as the node capacity of $s_1$ is varied. The relative-deadline was fixed at $3$ time-slots for flow packets of flow $f_1$, and at $2$ time slots for packets of flows $f_2$ and $f_3$. Each flow was assumed to receive $50$ packets per time slot. Reliability of the link $s_1 \to d_2$ was fixed at $.5$, while those of other links were assumed to be $1$.}
	\label{fig7}
\end{figure}

For the network shown in Figure~\ref{fig3}, the channel reliabilities of each link are fixed at $1$, except for the link $s_1 \to d_2$, which is fixed at $0.5$. Packets belonging to flow $f_1$ have a relative deadline of $3$ time slots, while packets belonging to flows $f_2$, and $f_3$ have a relative deadline of $2$ time slots. Moreover it is assumed that each flow receives $50$ packets in each time slot. The node capacities of each node except the node $s_1$ are assumed to be $100$ packets/slot. The throughputs of all the flows are weighted equally, i.e., $\alpha_1=\alpha_2=\alpha_3=1$. Figure~\ref{fig7} compares the throughput attained by the optimal policy with that attained by the EDF policy as a function of capacity of node $s_1$. 
\section{Conclusions}
We have posed the problem of designing throughput-maximizing scheduling policies for deadline-constrained multi-hop wireless networks as a Constrained Markov Decision Process (CMDP). Though a first look at the complex nature of the problem suggests that the optimal policy should make a scheduling decision based on the entire state of the network, and thus suggests that each node in the network must know the state of the network, we have however solved the dual of the original CMDP and thereby derived an optimal policy that is highly decentralized. In order to implement this optimal policy, each node in the network needs to know only the time-till-deadline of each packet present with it. 

Furthermore, the problem of obtaining the optimal Lagrange multipliers or the node prices has also been shown to have a decentralized solution, in which each node monitors its own congestion and updates the prices accordingly. The node-prices have been shown to be a means of congestion control. Simulations show that the optimal policy indeed out performs the Earliest Deadline First policy, comporting with the theory. 
\bibliographystyle{IEEEtran}
\bibliography{sigproc}
\end{document}